\documentclass[
    ,final            
  ]
  {aipproc}
\layoutstyle{8x11double}

\begin{document}

\title{Upflows in the upper transition region of the quiet Sun}



\author{H. Tian}{
  address={School of Earth and Space Sciences, Peking University, Beijing 100871, China}
  ,altaddress={Max-Planck-Institut f\"{u}r Sonnensystemforschung, 37191 Katlenburg-Lindau, Germany} 
}
\author{C.-Y. Tu}{
  address={School of Earth and Space Sciences, Peking University, Beijing 100871, China}
}
\author{E. Marsch}{
  address={Max-Planck-Institut f\"{u}r Sonnensystemforschung, 37191 Katlenburg-Lindau, Germany}
}
\author{J.-S. He}{
  address={Max-Planck-Institut f\"{u}r Sonnensystemforschung, 37191 Katlenburg-Lindau, Germany}
}
\author{C. Zhou}{
  address={Department of Atmospheric, Oceanic and Space Sciences, University of Michigan, USA}
}
\author{L. Zhao}{
  address={Department of Atmospheric, Oceanic and Space Sciences, University of Michigan, USA}
}

\begin{abstract}
 We investigate the physical meaning of the prominent blue shifts of Ne~{\sc{viii}},
 which is observed to be associated with quiet-Sun network junctions (boundary intersections),
 through data analyses combining force-free-field extrapolations
 with EUV spectroscopic observations. For a middle-latitude region, we reconstruct
 the magnetic funnel structure in a sub-region showing faint emission
 in EIT-Fe 195. This funnel appears to consist of several smaller funnels
 that originate from network lanes, expand with height and finally
 merge into a single wide open-field region. However, the large blue shifts
 of Ne~{\sc{viii}} are generally not associated with open fields,
 but seem to be associated with the legs of closed
 magnetic loops. Moreover, in most cases significant upflows are found in both
 of the funnel-shaped loop legs. These quasi-steady upflows are regarded as
 signatures of mass supply to the coronal loops rather than the solar wind.
 Our observational result also reveals that in many cases the upflows in the upper transition
 region (TR) and the downflows in the middle TR are not fully cospatial. Based
 on these new observational results, we suggest different TR structures in
 coronal holes and in the quiet Sun.
\end{abstract}

\maketitle


\section{Introduction}

It has well been established that the Ne~{\sc{viii}}~(770.4~{\AA})
line is on average blue shifted in coronal holes and the quiet Sun.
This line is formed in the upper transition region (TR) and lower
corona, and thus its Doppler shift is very useful to study TR
dynamics and the solar wind origin. In coronal holes (CHs), the blue
shift of Ne~{\sc{viii}} is believed to be an indicator of solar wind
outflow \citep[e.g.,][]{Hassler1999,Wilhelm2000,Tu2005a}. Also in
the quiet Sun, significant blue shifts of Ne~{\sc{viii}} were found
at the network junctions and interpreted \citep{Hassler1999} to be
possible sources of the solar wind. This idea challenges the
conventional notion that the solar wind originates from coronal
holes. To help clarifying this issue we have, through data analyses
combining force-free-field extrapolation with EUV spectroscopy,
further studied the prominent blue shifts of Ne~{\sc{viii}} in the
quiet Sun and investigated its physical meaning.

\section{Observational result}

The analyzed spectroscopic data were taken by SUMER (Solar
Ultraviolet Measurements of Emitted Radiation) onboard SOHO (Solar
and Heliospheric Observatory) in a middle-latitude quiet-Sun region,
during the time from 00:40 to 08:15 UTC on 22 September 1996. The details
of this observation can be found in \cite{Hassler1999}, \cite{Tu2005b},
\cite{Dammasch1999}, and \cite{Tian2008a}. A magnetogram taken by
MDI (Michelson Doppler Imager) at 01:39 UTC, and an EIT (Extreme
ultraviolet imaging telescope) image (Fe~{\sc{xii}} 19.5 nm) taken
at 01:06 UTC on the same day were also used in our study.

The linear force-free-field model equations as proposed by \cite{Seehafer1978}
were applied to construct the 3-D coronal magnetic field structure above the
photosphere in this quiet region. As suggested by previous work \citep{Tu2005b},
a simple potential field seems appropriate for our data. Therefore, we used
a coronal potential field.

\subsection{Relationship between upflows and magnetic funnels}

\begin{figure}
  \includegraphics[width=.45\textwidth]{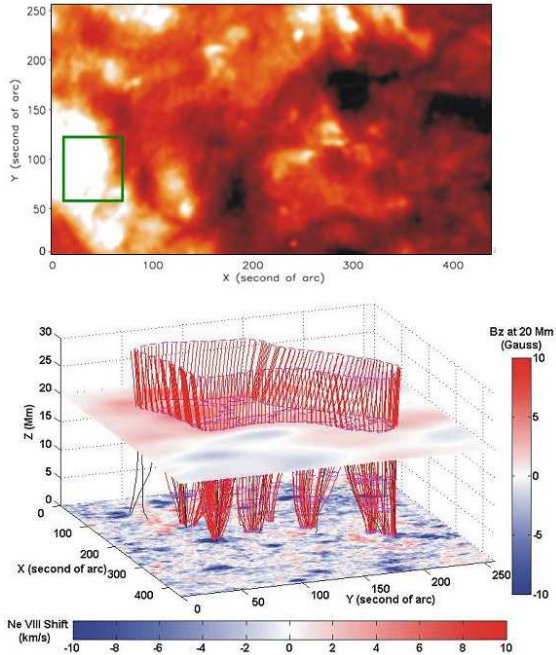}
  \caption{Upper panel: An EIT image (19.5 nm) taken
at 01:06 UTC. The rectangle outlines the region shown in
Fig.~\ref{fig.2}. Lower panel: Reconstructed magnetic funnels in the
observed region. The red lines indicate field lines originating from
the funnel boundary, and the black ones are open field lines outside
the small funnels. The Dopplergram of Ne~{\sc{viii}} is placed at
the bottom layer. The extrapolated longitudinal field strength at
20~Mm is also shown as a magnetogram placed at the corresponding
height. The figure is adapted from \cite{Tian2008a}. }
  \label{fig.1}
\end{figure}

The quiet corona as imaged by EIT at 19.5 nm and shown in
Fig.~\ref{fig.1} reveals some bright emission features, which are
known as EUV bright points, and some very dark emission patches. As
noted by \cite{He2007}, one dark region is associated with open
field lines. In \cite{Tian2008a} it was further pointed out that
these open field lines may correspond to coronal funnels.

The lower panel of Fig.~\ref{fig.1} shows the reconstructed magnetic
funnels in the observed region. The funnel structure consists of two
parts: below 20~Mm, there are several small funnels which originate
in the photosphere and expand with height; above 20~Mm, these small
funnels merge into a single wide open-field region. One may
speculate that coronal funnel structures in CHs should be of a
similar form.

However, a comparison of the Dopplergram of Ne~{\sc{viii}} with the
3-D funnel structures in fact reveals that most of the patches with
significant Ne~{\sc{viii}} blue shifts do not coincide with the
funnel legs. Thus, as argued before by \cite{Tian2008a}, these sites
might not be sources of the solar wind. On the other hand, it is
possible that in the quiet-Sun region the layer where the solar wind
outflow actually starts is located higher up than the source of the
Ne~{\sc{viii}} emission. If this is the case, significant outflows
could perhaps be found in the Dopplergram of another ion line that
forms at a higher temperature than that of Ne~{\sc{viii}}.

\subsection{Relationship between upflows and magnetic loops}

\begin{figure}
  \includegraphics[width=.45\textwidth]{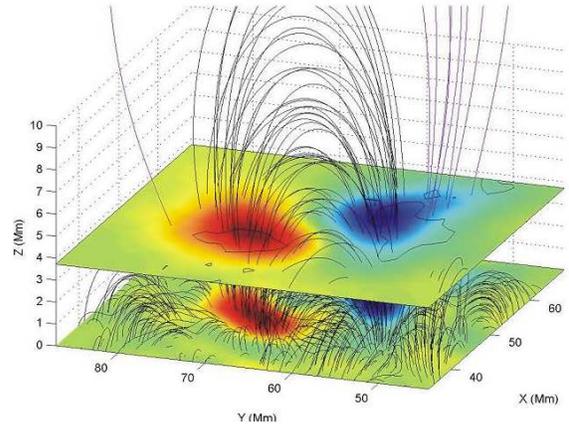}
  \caption{3-D illustration of magnetic loops. The extrapolated
  magnetograms at 0~Mm and 4~Mm are shown and placed at the corresponding
  heights. The strongest magnetic field with different
  polarities are shown in red and blue, respectively. The black contours
  outline patches with significant blue shift of Ne~{\sc{viii}} and encircle
  the strong-field regions.}
  \label{fig.2}
\end{figure}

So far we have argued that the quiet-Sun coronal funnels are not
associated with prominent outflows in the upper TR. On the contrary,
the relationship is very clear between significant outflow and the
related field lines being closed. As mentioned in \cite{Tian2008a},
there are several magnetic loops revealing upflows in both legs, and
some other loops with upflow in one yet downflow in the other leg.

More recently, the association of significant Ne~{\sc{viii}} blue
shift with the loop legs has been further confirmed. In a new
observation, \cite{Tian2009} found clearly that almost all of the
patches with a significant blue shift on the Dopplergram of
Ne~{\sc{viii}} coincided with legs of loops located at network
intersections. Furthermore, we found that most of the blue-shift
patches coincided with both legs of the magnetic loops, and some
patches were associated with the common leg of several joint loops.

The 2-D illustration of such a relationship can be found in
\cite{Tian2008a} and \cite{Tian2009}. Here we present a 3-D
illustration, which is given in Fig.~\ref{fig.2}. This reconstructed
loop system is associated with an EUV bright point located on the
eastern side of the FOV (field of view) shown in Fig.~\ref{fig.1}.
The figure clearly reveals that both legs of the loop system are
associated with upflows in the upper TR.

Since in the quiet Sun the blue shift of Ne~{\sc{viii}} has a closer
relationship with coronal loops than with open field lines, we
therefore suggest that the in the quiet Sun it is a signature of
mass supply to coronal loops rather than a signature of solar wind
origin.

\subsection{Spatial correlation between upflows in the upper TR and downflows
in the middle TR}

\begin{figure}
  \includegraphics[width=.45\textwidth]{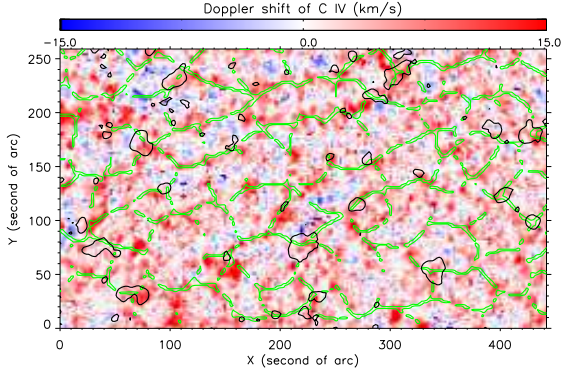}
  \caption{The contours (closed black lines) of strong Ne~{\sc{viii}} blue
  shift are superposed on a Dopplergram of C~{\sc{iv}}. The pattern
  of the chromospheric network is shown in green.}
  \label{fig.3}
\end{figure}

It has been well established that the EUV lines formed in the
(middle) TR are redshifted by a few km/s
\citep[e.g.,][]{PeterJudge1999,Xia2003,Curdt2008}. Since the
redshifts in the middle TR and the blueshifts in the upper TR are
both strongest in the network, their relationship should be studied
in detail. In the scenario of solar wind origin as suggested in
\cite{Tu2005a}, the simultaneously observed blue and red shifts were
interpreted as upflow and downflow caused by magnetic reconnection
between open field lines in coronal funnels and closed ones in
adjacent loops.

In the recent studies of \cite{Aiouaz2008} and \cite{Tian2008b}, the
spatial relationship between upflows in the upper TR and downflows
in the middle TR was investigated. Through a statistical analysis
\cite{Tian2008b}, it was concluded that the blue shift of
Ne~{\sc{viii}} tends to be strongest at the network center, while
the location of the maximum red shift of C~{\sc{iv}} seems to
deviate slightly from the network center.

Here we superpose the contours of strong Ne~{\sc{viii}} blue shifts
on the Dopplergram of C~{\sc{iv}}, and show them in
Fig.~\ref{fig.3}. The pattern of the chromospheric network, which
was extracted from a Si~{\sc{ii}} intensity image and reproduced
from the paper by \cite{Hassler1999}, is shown in green. This figure
reveals a clear near-cospatiality of strong up and down flows. We
further noticed that in many cases the prominent blue shifts do not
fully coincide with but slightly deviate from the strongest red
shifts of C~{\sc{iv}}. While in some other cases they strictly
coincide.

Different mechanisms such as cooling of spicules, siphon flows
through loops, downward propagation of MHD waves produced by
nano-flares, and downflows resulting from reconnection all can cause
red shifts of TR lines \citep{PeterJudge1999,Tu2005a}. If we assume
that those strong red shifts of C~{\sc{iv}} which are almost
cospatial with the prominent Ne~{\sc{viii}} blue shifts are mainly
caused by magnetic reconnection between legs of large loops and cool
side loops, then it is natural to see the near-cospatiality of
strong up and down flows at different layers. However, it is unclear
why in many cases the prominent blue shifts do not fully coincide
with the strongest red shifts of C~{\sc{iv}}.

\section{Magnetic structures of the transition region}

\begin{figure}
  \includegraphics[width=0.60\textwidth]{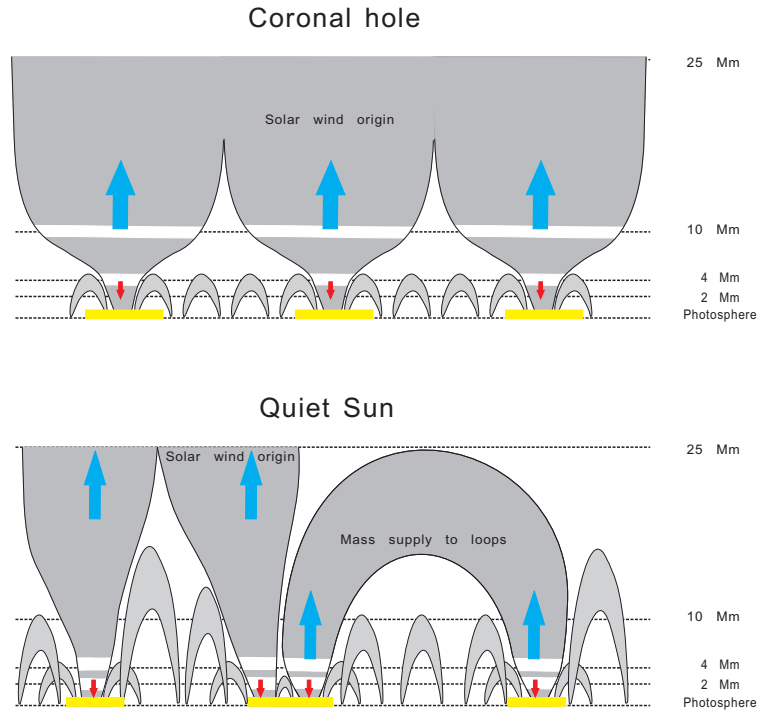}
  \caption{Schematic representation of the magnetic structures of
  the transition region in coronal holes (upper) and quiet Sun
  (lower). The top and bottom of the TR are indicated by the two white
  bars in each funnel or loop leg. The TR downflows and upflows in
  upper layers are marked in red and blue, respectively. The yellow
  bars indicate the supergranular boundaries.}
  \label{fig.4}
\end{figure}

Based on our recent observations of the quiet Sun and previous
observational results for coronal holes, we conclude that the transition
region consists of different magnetic structures in coronal holes than
quiet Sun, as they are sketched in Fig.~\ref{fig.4}.

In CHs the dominant magnetic structures are coronal funnels which
are connected to the solar wind \citep{Tu2005a}. Most of the
magnetic loops reside in the lower part of the atmosphere. There are
almost no large loops \citep{Wiegelmann2004,Zhang2006,Tian2008c}.
Thus, in CHs magnetic funnels can expand strongly in the TR
\citep{Tian2008c}. While magnetic loops of different spatial scales
are crowded in the quiet Sun, so that large coronal loops, whose
legs also exhibit funnel-like shapes \citep{Peter2001}, can not
drastically expand with height. Also, the dense loops around funnel
legs only permit a weaker expansion of possible quiet-Sun funnels.
Some blueshifts in the quiet-Sun network can possibly work their way
up to higher layers, finally merge and form the slow solar wind. A
similar scenario can also be found in \cite{He2009}.

The emission heights of TR lines are also different in the two
regions. Based on the results in \cite{Tu2005a}, \cite{Tu2005b}, and
\cite{Tian2008d}, the height extension of the TR is approximately
4-10~Mm in coronal holes and only 2-4~Mm in the quiet Sun. However, we
have to mention that this height range is only a rough estimate, since the
TR is known to be strongly non-uniform \citep{Marsch2006}. Let us consider
finally these results in a wider context.

The concept of coronal circulation as proposed in
\citep{Marsch2008}, or coronal convection, emphasizes that the
coronal plasma is nowhere static but observed to be flowing
everywhere and guided by various magnetic channels. The various
kinds of flows at large scales all appear long lasting, and thus may
indicate quasi-steady plasma circulation in the entire corona and
TR. The upflows in the upper TR and downflows in the middle TR
network intersections are also observed to be long lasting, and thus
should play a role in this process. In coronal holes, restless
magnetic reconnection between the field lines in coronal funnels and
their side loops is likely to supply persistently mass and energy to
what becomes a fast solar wind stream. In the quiet Sun, as
mentioned previously, continuous magnetic reconnection is likely to
be the mechanism responsible for the bi-directional flows in the TR.
On the other hand, if cool plasma continuously enters any loop leg
through a certain process from outside, a heating process might
cause its upflow into the upper TR. At a certain height below the
apex of the loop, the flow might, possibly due to cooling, turn
downwards again.

We have to point out that our main results are based on the
assumption of a force-free field. As mentioned in \cite{Tu2005b},
the force-free model works well for long-living structures, and the
results are not severely influenced by the force-free parameter.
Moreover, the excellent agreement between the EUV emission pattern
of a coronal bright point and the extrapolated loop structure also
indicates the validity of our extrapolation model \citep{Tian2009}.

\begin{theacknowledgments}
The SUMER project is financially supported by DLR, CNES, NASA, and
the ESA PRODEX programme (Swiss contribution). SUMER and MDI are
instruments on board SOHO, an ESA and NASA mission. The work of Hui
Tian and Chuanyi Tu at Peking University is supported by NSFC under
contract 40874090.

\end{theacknowledgments}

\end{document}